\documentstyle{l-aa}

\def\aj#1{{\it Astron. J.,} {\bf #1}}
\def\apj#1{{\it Astrophys. J.,} {\bf #1}}
\def\apjss#1{{\it Astrophys. J. Suppl. Ser.,} {\bf #1}}
\def\aa#1{{\it Astron. \& Astrophys.,} {\bf #1}}
\def\araa#1{{\it Annu. Rev. Astron. \& Astrophys.,} {\bf #1}}

\def\aas#1{{\it Astron. \& Astrophys. Suppl.,} {\bf #1}}
\def\mn#1{{\it Mon. Not. Roy. Astr. Soc.,} {\bf #1}}
\def\kfnt#1#2{{\it Kinematika i Fizika Nebesnykh Tel,} {\bf #1}, {No. #2}}
\def\ass#1{{\it Astrophys. \& Sp. Sci.,} {\bf #1}}

\def\be{\begin{equation}}
\def\ee{\end{equation}}

\def\bea{\begin{eqnarray}}
\def\eea{\end{eqnarray}}

\def\eg{{\it e.g.}}
\def\ie{{\it i.e.}}
\def\etal{{\it et al.}}

\def\citeauthoryear#1#2{{#1}{ #2}}

\def\vsp{\vspace{6.5cm}}

\def\insertplot#1{
\includegraphics{#1}}

\def\insplt#1{
\includegraphics{#1} }

\begin{document}

\thesaurus{11        
          (11.01.1;  
           11.05.2;  
           11.07.1;  
           11.11.1;  
           11.19.6)} 

\title{Chemo--Dynamical SPH code for evolution of star forming disk galaxies.}


\author{Peter Berczik}

\offprints{Peter Berczik}

\institute{Main Astronomical Observatory of
           Ukrainian National Academy of Sciences,
           252650, Golosiiv, Kiev-022, Ukraine \\
           e-mail: {\tt berczik@mao.kiev.ua}}

\date{Received: 16 February, 1999 / Accepted: 30 April, 1999}

\maketitle


\begin{abstract}

A new Chemo--Dynamical Smoothed Particle Hydrodynamic (CD-SPH) code is
presented. The disk galaxy is described as a multi--fragmented gas and
star system, embedded in a cold dark matter halo with a rigid potential
field. The star formation (SF) process, SNII, SNIa and PN events, and
the chemical enrichment of gas, have all been considered within the
framework of the standard SPH model, which we use to describe the
dynamical and chemical evolution of triaxial disk--like galaxies. It is
found that such approach provides a realistic description of the
process of formation, chemical and dynamical evolution of disk galaxies
over a cosmological timescale.


\end{abstract}


\section{Introduction}

The dynamical and chemical evolution of galaxies is one of the most
interesting and complex of problems. Naturally, galaxy formation is
closely connected with the process of large--scale structure formation
in the Universe.

The main role in the scenario of large--scale structure formation seems
to be played by the dark matter. It is believed that the Universe was
seeded at some early epoch with low density fluctuations of dark
non--baryonic matter, and the evolving distribution of these dark halos
provides the arena for galaxy formation. Galaxy formation itself
involves collapse of baryons within potential wells of dark halos
(\cite{WR78}). The properties of forming galaxies depend on the amount
of baryonic matter that can be accumulated in such halos and the
efficiency of star formation. The observational support for this galaxy
formation scenario comes from the recent COBE detection of fluctuations
in the microwave background (\eg~ \cite{BBHMSW93}).

The investigation of galaxy formation is a highly complex subject
requiring many different approaches. The formation of self--gravitating
inhomogeneities of protogalactic size, the ratio of baryonic and
non--baryonic matter (\cite{BBKSz86}; \cite{WS79}; \cite{P93};
\cite{D95}), the origin of the protogalaxy's initial angular momentum
(\cite{VH89}; \cite{ZQS88}; \cite{EL95}; \cite{SB95}), and the
protogalaxy's collapse and its subsequent evolution are all usually
considered as separate problems. Recent advances in computer technology
and numerical methods have allowed detailed modeling of baryon matter
dynamics in the universe dominated by collisionless dark matter and,
therefore, the detailed gravitational and hydrodynamical description of
galaxy formation and evolution. The most sophisticated models include
radiative processes, star formation and supernova feedback (\eg~
\cite{K92}; \cite{SM94}; \cite{FB95}).

The results of numerical simulations are essentially affected by the
star formation algorithm incorporated into modeling techniques. The
star formation and related processes are still not well understood on
either small or large spatial scales, such that the star formation
algorithm by which the gas material is converted into stars can only be
based on simple theoretical assumptions or on empirical observations of
nearby galaxies. The other most important effect of star formation on
the global evolution of a galaxy is caused by a large amount of energy
released in supernova explosions and stellar winds.

Among numerous methods developed for the modeling of complex three
dimensional hydrodynamic phenomena, Smoothed Particle Hydrodynamics
(SPH) is one of the most popular (\cite{M92}). Its Lagrangian nature
allows easy combination with fast N--body algorithms, making it
suitable for simultaneous description of the complex dynamics of a
gas--stellar system (\cite{FB95}). As an example of such a combination,
the TREE-SPH code (\cite{HK89}; \cite{NW93}) was successfully applied
to the detailed modeling of disk galaxy mergers (\cite{MH96}) and of
galaxy formation and evolution (\cite{K92}). The second good example is
a GRAPE-SPH code (\cite{SM94}; \cite{SM95}) which was successfully used
to model the evolution of disk galaxy structure and kinematics.

In recent years, there have been excellent papers concerning the
complex SPH modeling of galaxy formation and evolution (\cite{RVN96};
\cite{CLC97}). Our code, proposes new "energetic" criteria for SF, and
suggest a more realistic account of returned chemically enriched gas
fraction via SNII, SNIa and PN events.

The simplicity and numerical efficiency of the SPH method were the main
reasons why we chose this technique for the modeling of the evolution
of complex, multi--fragmented triaxial protogalactic systems. We used
our own modification of the hybrid N--body/SPH method (\cite{BerK96};
\cite{Ber98}), which we call the chemo--dynamical SPH (CD-SPH) code.

The "dark matter" and "stars" are included in the standard SPH
algorithm as the N--body collisionless system of particles, which can
interact with the gas component only through gravitation (\cite{K92}).
The star formation process and supernova explosions are also included
in the scheme as proposed by
Raiteri \etal~ (1996), but with our own modifications.


\section{The CD-SPH code}

\subsection{The SPH code}

Continuous hydrodynamic fields in SPH are described by the
interpolation functions constructed from the known values of these
functions at randomly positioned particles (\cite{M92}). Following
Monaghan \& Lattanzio (1985) we use for the kernel function $ W_{ij} $
the spline expression in the form of:

\bea
   W_{ij} = \frac{1}{\pi h^3}
            \left\{
            \begin{array}{lllll}
            1 - \frac{3}{2} u_{ij}^2 + \frac{3}{4} u_{ij}^3, & \mbox{ if $0 \leq u_{ij} < 1$}, \\
                                                                                                            \\
            \frac{1}{4} (2  - u_{ij})^3,                     & \mbox{ if $1 \leq u_{ij} < 2$}, \\
                                                                                                            \\
            0,                                               & \mbox{ otherwise }. \\
            \end{array}
            \right.
   \label{eq:def_w}
\eea

     Here $ u_{ij} = r_{ij}/h $.

To achieve the same level of accuracy for all points in the fluid, it
is necessary to use a spatially variable smoothing length. In this case
each particle has its individual value of $ h $. Following
Hernquist \& Katz (1989), we write:

\be
   < \rho({\bf r}_i) > = \sum_{j=1}^{N} m_j \cdot \frac{1}{2} \cdot
   [W(r_{ij}; h_i) + W(r_{ij}; h_j)].
   \label{eq:def_rho_new}
\ee

In our calculations the values of $ h_i $ were determined from the
condition that the number of particles $ N_B $ in the neighborhood of
each particle within the $ 2 \cdot h_i $ remains constant
(\cite{MH96}). The value of $ N_B $ is chosen such that a certain
fraction of the total number of "gas" particles $ N $ affects the local
flow characteristics (\cite{HV91}). If the defined $ h_i $ becomes
smaller than the minimal smoothing length $ h_{min} $, we set the value
$ h_i = h_{min} $. For "dark matter" and "star" particles (with Plummer
density profiles) we use, accordingly, the fixed gravitational
smoothing lengths $ h_{dm} $ and $ h_{star} $.

If  the density is computed according to Equation
(\ref{eq:def_rho_new}), then the continuity equation is satisfied
automatically. Equations of motion for particle $ i $ are

\be
   \frac{d{\bf r}_i}{dt} = {\bf v}_i,
   \label{eq:def_r}
\ee

\be
   \frac{d{\bf v}_i}{dt} = -\frac{\nabla_i P_i}{\rho_i} +
                    {\bf a}^{vis}_{i} -
                    \nabla_i \Phi_i -
                    \nabla_i \Phi^{ext}_i,
   \label{eq:def_v}
\ee

where $ P_i $ is the pressure, $ \Phi_i $ is the self gravitational
potential, $ \Phi^{ext}_i $ is a gravitational potential of possible
external halo and $ {\bf a}^{vis}_{i} $ is an artificial viscosity term
(\cite{HVC91}). The energy equation has the form:

\be
   \frac{du_i}{dt} = -\frac{P_i}{\rho_i} \nabla_i {\bf v}_i +
                      \frac{\Gamma_i - \Lambda_i}{\rho_i}.
   \label{eq:def_du}
\ee

Here $ u_i $ is the specific internal energy of particle $ i $. The
term  $ (\Gamma_i - \Lambda_i)/\rho_i $ accounts for non adiabatic
processes not associated with the artificial viscosity (in our
calculations $ \Gamma_i \equiv 0 $). We present the radiative cooling
in the form:

\be
   \Lambda_i = \Lambda_i(u_i, \rho_i) = \Lambda^{*}_i(T_i) \cdot n^2_i,
   \label{eq:lam_i}
\ee

where $ n_i $ is the hydrogen number density and $ T_i $ the
temperature. To follow its subsequent thermal behaviour in numerical
simulations, we use an analytical approximation of the standard cooling
function $ \Lambda^{*}(T) $ for an optically thin primordial plasma in
ionization equilibrium (\cite{DM72}; \cite{KG91}). Its absolute cutoff
temperature is set equal to $ 10^4 $ K.

The equation of state must be added to close the system.

\be
   P_i = \rho_i \cdot (\gamma - 1) \cdot u_i,
   \label{eq:def_P_i_1}
\ee

where $ \gamma = 5/3 $ is the adiabatic index.


\subsection{Time integration}

To solve the system of Equations (\ref{eq:def_r}), (\ref{eq:def_v}) and
(\ref{eq:def_du}) we use the leapfrog integrator (\cite{HK89}). The
time step $ \delta t_i $ for each particle depends on the particle's
acceleration $ {\bf a}_i $ and velocity $ {\bf v}_i $, as well as on
viscous forces. To define $ \delta t_i $ we use the relation from
Hiotelis \& Voglis (1991), and adopt Courant's number $ C_n = 0.1 $.

We carried out (\cite{BerK93}) a large series of test calculations to
check that the code is correct, the conservation laws are obeyed and
the hydrodynamic fields are represented adequately, all with good
results.


\subsection{The star formation algorithm}

It is well known that star formation (SF) regions are associated with
giant molecular complexes, especially with regions that are approaching
dynamical instability. The early phase of star formation does not seem
to crucially affect the dynamics of a galaxy. From the beginning of the
collapse, such a system decouples from its surroundings and evolves on
a completely different timescale. When the chemically enriched gas
content of the galaxy decreases, the heating by winds and supernova
explosions (\cite{LRD92}) begins to play an important role in the
dynamics of the galaxy. The overall picture of star formation seems to
be understood, but the detailed physics of star formation and
accompanying processes, on either small or large scales, remains
sketchy (\cite{L69}; \cite{S87}).

All the above stated as well as computer constrains cause the using of
simplified numerical algorithms of description of conversion of the
gaseous material into stars, which are based on simple theoretical
assumptions and/or on results of observations of nearby galaxies.

To describe of the process of converting of gaseous material into stars
we modify the standard SPH star formation algorithm (\cite{K92};
\cite{NW93}), taking into account the presence of random motions in the
gaseous environment and the time lag between the initial development of
suitable conditions for star formation and star formation itself
(\cite{BerK96}; \cite{Ber98}). The first reasonable requirement
incorporated into this algorithm allows selecting "gas" particles that
are potentially eligible to form stars. It states that in the separate
"gas" particle the SF can start if the absolute value of the "gas"
particle's gravitational energy exceeds the sum of its thermal energy
and the energy of random motions:

\be
   \mid E_i^{gr} \mid > E_i^{th} + E_i^{ch}.
  \label{eq:crit}
\ee

Gravitational and thermal energies and the energy of random motions for
the "gas" particle $ i $ in model simulation are defined as:

\bea
  \left\{
  \begin{array}{lllll}
  E_i^{gr} = - \frac{3}{5} \cdot G \cdot m_i^2/h_i,    \\
                                                       \\
  E_i^{th} = \frac{3}{2} \cdot m_i \cdot c_i^2,        \\
                                                       \\
  E_i^{ch} = \frac{1}{2} \cdot m_i \cdot \Delta v_i^2, \\
  \end{array}
  \right.
  \label{eq:energy}
\eea

where $ c_i = \sqrt{\Re \cdot T_i / \mu} $ is the isothermal sound
speed of particle $ i $. We set $ \mu = 1.3 $ and define the random or
"turbulent" square velocities near particle $ i $ as:

\be
  \Delta v_i^2 = \sum_{j=1}^{N_{B}} m_j \cdot ({\bf v}_j - {\bf v}_c)^2 /
                 \sum_{j=1}^{N_{B}} m_j,
\ee

where:

\be
  {\bf v}_c = \sum_{j=1}^{N_{B}} m_j \cdot {\bf v}_j /
              \sum_{j=1}^{N_{B}} m_j.
\ee

For practical reasons, it is useful to define a critical temperature
for SF onset in particle $ i $ as:

\be
  T^{crit}_i = \frac{\mu}{3 \Re} \cdot
               ( \frac{8}{5} \cdot \pi \cdot G \cdot \rho_i \cdot h_i^2
               - \Delta v_i^2 ).
  \label{eq:t_crit}
\ee

Then, if the temperature of the "gas" particle $ i $, drops below the
critical one, SF can proceed.

\be
  T_i < T^{crit}_i.
  \label{eq:crit_T}
\ee

We think that requirement (\ref{eq:crit}), or in another form
(\ref{eq:crit_T}), is the only one needed. It seems reasonable that the
chosen "gas" particle will produce stars only if the above condition
hold over the interval that exceeds its free - fall time $ t_{ff} =
\sqrt {3 \cdot \pi / ( 32 \cdot G \cdot \rho ) } $. This condition is
based on the well known fact that, due to gravitational instability,
all substructures of a collapsing system are formed on such a
timescale. Using it, we exclude transient structures, that are
destroyed by the tidal action of surrounding matter from consideration.

We also define which "gas" particles remain cool, \ie~ $ t_{cool} <
t_{ff} $. We rewrite this condition as presented in
Navarro \& White (1993): $ \rho_i > \rho_{crit} $. Here we use the
value of $ \rho_{crit} = 0.03 $ cm$^{-3}$.

When the collapsing particle $ i $ is defined, we create the new "star"
particle with mass $ m^{star} $ and update the "gas" particle $ m_i $
using these simple equations:

\bea
  \left\{
  \begin{array}{lll}
  m^{star} = \epsilon \cdot m_i,  \\
                                  \\
  m_i = (1 - \epsilon) \cdot m_i. \\
  \end{array}
  \right.
  \label{eq:m}
\eea

Here $ \epsilon $, defined as the global efficiency of star formation,
is the fraction of gas converted into stars according to the
appropriate initial mass function (IMF). The typical values for SF
efficiency in our Galaxy on the scale of giant molecular clouds are in
the range $ \epsilon \approx 0.01 \div 0.4 $ (\cite{DIL82};
\cite{WL83}). But it is still a little known quantity. In numerical
simulation the model parameter has to be checked by comparison of
numerical simulation results with available observational data. Here we
define $ \epsilon $ as:

\be
  \epsilon = 1 - (E_i^{th} + E_i^{ch})/\mid E_i^{gr} \mid,
  \label{eq:epsilon}
\ee

with the requirement that all excess mass of the gas component in a
star--forming particle, which provides the inequality $ \mid E_i^{gr}
\mid > E_i^{th} + E_i^{ch} $, is transformed into the star component.
In the code we set the absolute maximum value of the mass of such a
"star" particle $ m^{star}_{max} = 2.5 \; 10^6 \; M_\odot $ \ie~ $
\approx 5 \% $ of the initial particle mass $ m_i $.

At the moment of stellar birth, the position and velocities of new
"star" particles are equal to those of parent "gas" particles.
Thereafter these "star" particles interact with other "gas", "star" or
"dark matter" particles only by gravitation. The gravitational
smoothing length for these (Plummer like) particles is set equal to $
h_{star} $.


\subsection{The thermal SNII feed--back}

We try to include the events of SNII, SNIa and PN in the complex
gasdynamic picture of galaxy evolution. But, for the thermal budget of
the ISM, only SNII plays the main role. Following
Katz (1992), we assume that the explosion energy is converted totally
into thermal energy. The stellar wind action seems not to be essential
in the energy budget (\cite{F95}). The total energy released by SNII
explosions ($ 10^{44} $ J per SNII) within a "star" particle is
calculated at each time step and distributed uniformly between the
surrounding (\ie~ $ r_{ij} < h_{star} $) "gas" particles
(\cite{RVN96}).


\subsection{The chemical enrichment of gas}

Every "star" particle in our SF scheme represents a separate,
gravitationally closed star formation macro region (like a globular
cluster). The "star" particle is characterized by its own time of birth
$ t_{begSF} $ which is set equal to the moment of particle formation.
After the formation, these particles return the chemically enriched gas
into surrounding "gas" particles due to SNII, SNIa and PN events. For
the description of this process we use the approximation proposed by
Raiteri \etal~ (1996). We consider only the production of $^{16}$O and
$^{56}$Fe, and try to describe the full galactic time evolution of
these elements, from the beginning to present time (\ie~ $ t_{evol}
\approx 13.0 $ Gyr).

With the multi--power IMF law suggested by
Kroupa \etal~ (1993), the distribution of stellar masses within a
"star" particle of mass $ m^{star} $ is then:

\bea
   \Psi(m) = m^{star} \cdot A \cdot
            \left\{
            \begin{array}{lllll}
            2^{0.9} \cdot m^{-1.3}, & \mbox{ if $ 0.1  \leq m < 0.5 $}, \\
                                                                        \\
            m^{-2.2},               & \mbox{ if $ 0.5  \leq m < 1.0 $}, \\
                                                                        \\
            m^{-2.7},               & \mbox{ if $ 1.0  \leq m $},       \\
            \end{array}
            \right.
   \label{eq:def_Psi}
\eea

where $ m $ is the star mass in solar units. With adopted lower ($
m_{low} = 0.1 \; M_\odot $) and upper ($ m_{upp} = 100 \; M_\odot $)
limits of the IMF, the normalization constant $ A \approx 0.31 $.

For the definition of stellar lifetimes we use the equation
(\cite{RVN96}):

\be
  \log t_{dead} = a_0({\rm Z}) - a_1({\rm Z}) \cdot \log m + a_2({\rm Z}) \cdot (\log m)^2,
  \label{eq:def_tdead}
\ee

where $ t_{dead} $ is expressed in years, $ m $ is in solar units, and
coefficients are defined as:

\bea
  \left\{
  \begin{array}{lllll}
  a_0({\rm Z}) = 10.130 + 0.0755 \cdot \log {\rm Z} - 0.0081 \cdot (\log {\rm Z})^2, \\
                                  \\
  a_1({\rm Z}) = 4.4240 + 0.7939 \cdot \log {\rm Z} + 0.1187 \cdot (\log {\rm Z})^2, \\
                                  \\
  a_2({\rm Z}) = 1.2620 + 0.3385 \cdot \log {\rm Z} + 0.0542 \cdot (\log {\rm Z})^2. \\
  \end{array}
  \right.
  \label{eq:aZ}
\eea

These relations are based on the calculations of the Padova group
(\cite{ABB93}; \cite{BFBC93}; \cite{BBCFN94}) and give a reasonable
approximation to stellar lifetimes in the mass range from $ 0.6 \;
M_\odot $ to $ 120 \; M_\odot $ and metallicities Z $ = 7 \cdot 10^{-5}
\div 0.03 $ (defined as a mass of all elements heavier than He). In our
calculation following
Raiteri \etal~ (1996), we assume that Z scales with the oxygen
abundance as Z/Z$_\odot = ^{16}$O$/^{16}$O$_\odot$. For those
metallicities exceeding available data we take the value corresponding
to the extremes.

We can define the number of SNII explosions inside a given "star"
particle during the time from $ t $ to $ t + \Delta t $ using a simple
equation:

\be
  \Delta N_{{\rm SNII}} = \int\limits_{m_{dead}(t + \Delta t)}^{m_{dead}(t)} \Psi(m) dm,
  \label{eq:def_NSNII}
\ee

where $ m_{dead}(t) $ and $ m_{dead}(t + \Delta t) $ are masses of
stars that end their lifetimes at the beginning and at the end of the
respective time step. We assume that all stars with masses between $ 8
\; M_\odot $ and $ 100 \; M_\odot $ produce SNII, for which we use the
yields from
Woosley \& Weaver (1995). The approximation formulae from
Raiteri \etal~ (1996) defines the total ejected mass by one SNII - $
m^{tot}_{ej} $, as well as the ejected mass of iron - $ m^{{\rm
Fe}}_{ej} $ and oxygen - $ m^{{\rm O}}_{ej} $ as a function of stellar
mass (in solar units).

\bea
  \left\{
  \begin{array}{lllll}
  m^{tot}_{ej}       = 7.682 \cdot 10^{-1} \cdot m^{1.056}, \\
  \\
  m^{{\rm Fe}}_{ej}  = 2.802 \cdot 10^{-4} \cdot m^{1.864}, \\
  \\
  m^{{\rm O}}_{ej}   = 4.586 \cdot 10^{-4} \cdot m^{2.721}. \\
  \end{array}
  \right.
  \label{eq:m_SNII}
\eea

To take into account PN events inside the "star" particle we use the
equation, as for (\ref{eq:def_NSNII}):

\be
  \Delta N_{{\rm PN}} = \int\limits_{m_{dead}(t + \Delta t)}^{m_{dead}(t)} \Psi(m) dm.
  \label{eq:def_NPN}
\ee

Following
van den Hoek \& Groenewegen (1997), Samland (1997) and Samland \etal~
(1997), we assume that all stars with masses between $ 1 \; M_\odot $
and $ 8 \; M_\odot $ produce PN. We define the average ejected masses
(in solar units) of one PN event as (\cite{RV81}; \cite{vdHG97}):

\bea
  \left\{
  \begin{array}{lllll}
  m^{tot}_{ej}      = 1.63,  \\
                             \\
  m^{{\rm Fe}}_{ej} = 0.00,  \\
                             \\
  m^{{\rm O}}_{ej}  = 0.00.  \\
  \end{array}
  \right.
  \label{eq:m_PN}
\eea

The method described in
Raiteri \etal~ (1996) and proposed in
Greggio \& Renzini (1983) and Matteuchi \& Greggio (1986) is used to
account for SNIa. In simulations, the number of SNIa exploding inside a
selected "star" particle during each time step is given by:

\be
  \Delta N_{{\rm SNIa}} = \int\limits_{m_{dead}(t + \Delta t)}^{m_{dead}(t)} \Psi_2(m_2) dm_2.
  \label{eq:def_NSNIa}
\ee

The quantity $ \Psi_2(m_2) $ represents the initial mass function of
the secondary component and includes the distribution function of the
secondary's mass relative to the total mass of the binary system $
m_{B} $,

\be
  \Psi_2(m_2) = m^{star} \cdot A_2 \cdot
  \int\limits_{m_{inf}}^{m_{sup}} (\frac{m_2}{m_B})^2 \cdot m_B^{-2.7} dm_B,
  \label{eq:def_Psi2}
\ee

where $ m_{inf} = \max(2 \cdot m_2, \; 3 \; M_\odot) $ and $ m_{sup} =
m_2 + 8 \; M_\odot $. Following
van den Berg \& McClure (1994) the value of normalization constant we
set, equal to $ A_2 = 0.16 \cdot A $.

The total ejected mass (in solar units) is (\cite{TNY86};
\cite{NTY84}):

\bea
  \left\{
  \begin{array}{lllll}
  m^{tot}_{ej}      = 1.41, \\
                            \\
  m^{{\rm Fe}}_{ej} = 0.63, \\
                            \\
  m^{{\rm O}}_{ej}  = 0.13. \\
  \end{array}
  \right.
  \label{eq:m_SNIa}
\eea

In summary, a new "star" particle (with metallicity Z $ = 10^{-4} $)
with mass $ 10^4 \; M_\odot $ during the total time of evolution $
t_{evol} $ produces:

$$ \Delta N_{{\rm SNII}} \approx 52.5, \; \Delta N_{{\rm PN}} \approx
1770.0, \; \Delta N_{{\rm SNIa}} \approx 8.48. $$

\begin{figure}[htbp]

\vsp \insertplot{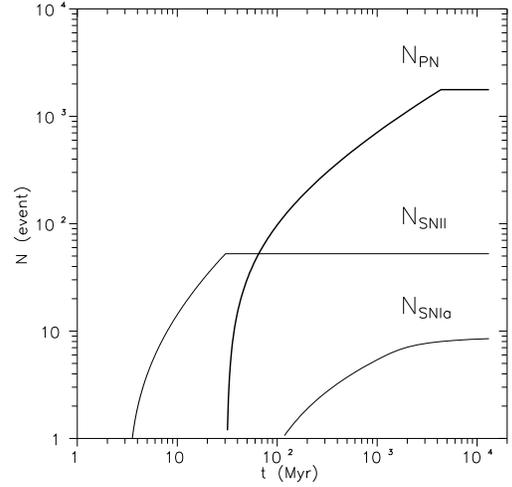}

\caption{The number of SNII, SNIa and PN events}

\label{fig-n}
\end{figure}

Fig.~\ref{fig-n}. presents the number of SNII, SNIa and PN events for
this "star" particle.

\begin{figure}[htbp]

\vsp \insertplot{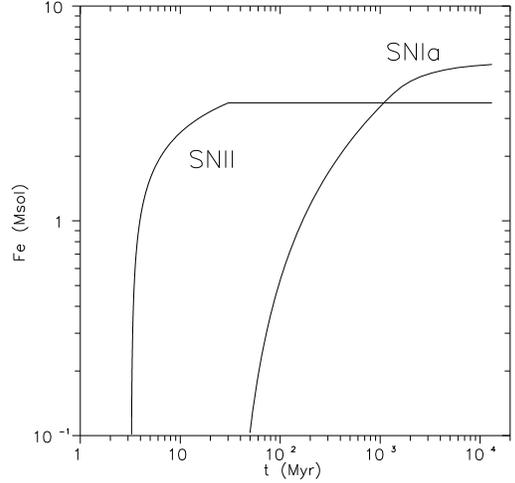}

\caption{The returned mass of $^{56}$Fe}

\label{fig-fe}
\end{figure}

\begin{figure}[htbp]

\vsp \insertplot{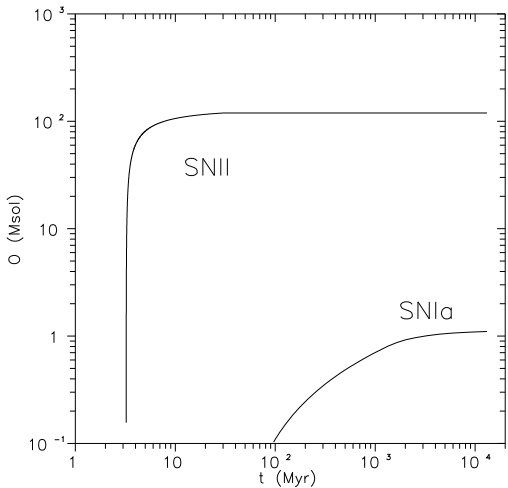}

\caption{The returned mass of $^{16}$O}

\label{fig-o}
\end{figure}

In Fig.~\ref{fig-fe}. and Fig.~\ref{fig-o}. we present the returned
masses of $^{56}$Fe and $^{16}$O. We can estimate the total masses (H,
He, $^{56}$Fe, $^{16}$O) (in solar masses) returned to the surrounding
"gas" particles due to these processes as:

\bea
  \left\{
  \begin{array}{lllllll}
  \Delta m^{{\rm H}}_{{\rm SNII}}  = 477, \;\;\; \Delta m^{{\rm H}}_{{\rm PN}}  = 2164,\;\;\;\; \Delta m^{{\rm H}}_{{\rm SNIa}}  = 4.14, \\
  \\
  \Delta m^{{\rm He}}_{{\rm SNII}} = 159, \;\;\; \Delta m^{{\rm He}}_{{\rm PN}} = 721.3, \;\;\; \Delta m^{{\rm He}}_{{\rm SNIa}} = 1.38, \\
  \\
  \Delta m^{{\rm Fe}}_{{\rm SNII}} = 3.5, \;\;\; \Delta m^{{\rm Fe}}_{{\rm PN}} = 0.000, \;\;\; \Delta m^{{\rm Fe}}_{{\rm SNIa}} = 5.35, \\
  \\
  \Delta m^{{\rm O}}_{{\rm SNII}}  = 119, \;\;\; \Delta m^{{\rm O}}_{{\rm PN}}  = 0.000, \;\;\; \Delta m^{{\rm O}}_{{\rm SNIa}}  = 1.10. \\
  \end{array}
  \right.
  \label{eq:masses}
\eea

\subsection{The cold dark matter halo}

In the literature we have found some profiles, sometimes controversial,
for the galactic Cold Dark Matter Haloes (CDMH) (\cite{B95};
\cite{N98}). For resolved structures of CDMH: $ \rho_{halo}(r) \sim
r^{-1.4} $ (\cite{MGQSL97}). The structure of CDMH high--resolution
N--body simulations can be described by: $ \rho_{halo}(r) \sim r^{-1} $
(\cite{NFW96}; \cite{NFW97}). Finally, in
Kravtsov \etal~ (1997), we find that the cores of DM dominated galaxies
may have a central profile: $ \rho_{halo}(r) \sim r^{-0.2} $.

In our calculations and as a first approximation, it is assumed that
the model galaxy halo contains the CDMH component with Plummer - type
density profiles (\cite{DC95}):

\be
\rho_{halo}(r) = \frac{ M_{halo} }{ \frac{4}{3} \pi b_{halo}^3 } \cdot
                 \frac{ b_{halo}^5 }{ (r^2 + b_{halo}^2 )^{\frac{5}{2}} },
\ee

Therefore for the external force acting on the "gas" and "star"
particles we can write:

\be
- \nabla_i \Phi^{ext}_i = - G \cdot \frac{ M_{halo} }
                            { ({\bf r}^2_{i} + b^2_{halo})^{\frac{3}{2}} }
                            \cdot {\bf r}_{i}.
\ee


\section{Results and discussion}

\subsection{Initial conditions}

After testing our code demonstrated that simple assumptions can lead to
a reasonable model of a galaxy. The SPH calculations were carried out
for  $ N_{gas} = 2109 $ "gas" particles. According to
Navarro \& White (1993) and Raiteri \etal~ (1996), such a number seems
adequate for a qualitatively correct description of the systems
behaviour. Even this small a number of "gas" particles produces $
N_{star} = 31631 $ "star" particles at the end of the calculation.

The value of the smoothing length $ h_i $ was chosen to require that
each "gas" particle had $ N_B = 21 $ neighbors within $ 2 \cdot h_i $.
Minimal $ h_{min} $ was set equal to $ 1 $ kpc, and the fixed
gravitational smoothing length $ h_{star} = 1 $ kpc was used for the
"star" particles. Our results show that a value of $ N_B \approx 1 \%
\; N_{gas} $ provides qualitatively correct treatment of the system's
large scale evolution.

As the initial model (relevant for CDM - scenario) we took a
constant--density homogeneous gaseous triaxial configuration ($ M_{gas}
= 10^{11} \; M_\odot $) within the dark matter halo ($ M_{halo} =
10^{12} \; M_\odot $). We set $ A = 100 $ kpc, $ B = 75 $ kpc and $ C =
50 $ kpc for semiaxes of system. Such triaxial configurations are
reported in cosmological simulations of the dark matter halo formation
(\cite{EL95}; \cite{FWDE88}; \cite{WQSZ92}). Initially, the centers of
all particles were placed on a homogeneous grid inside this triaxial
configuration. We set the smoothing parameter of CDMH: $ b_{halo} = 25
$ kpc. These values of $ M_{halo} $ and $ b_{halo} $ are typical for
CDMH in disk galaxies (\cite{NFW96}; \cite{NFW97}; \cite{B95}).

The gas component was assumed to be cold initially, $ T_0 = 10^{4} $ K.
As we see in our calculations, the influence of random motions
essentially reduces the dependence of model parameters on the adopted
temperature cutoff and, therefore, on the adopted form of the cooling
function itself.

The gas was assumed to be involved in the Hubble flow ($ H_{0} = 65 $
km/s/Mpc) and the solid - body rotation around $ z $ - axis. We added
small random velocity components  ($ \Delta \mid {\bf v} \mid = 10 $
km/s) to account for the random motions of fragments. The initial
velocity field was defined as:

\be
   {\bf v}(x, y, z) = [{\bf \Omega}(x, y, z) \times {\bf r}]
                         + H_{0} \cdot {\bf r}
                         + \Delta {\bf v}(x, y, z),
   \label{eq:V_0}
\ee

where $ {\bf \Omega}(x,y, z) $ is the angular velocity of an initially
rigidly rotating system.

The spin parameter in our numerical simulations is $ \lambda \approx
0.08 $, defined in
Peebles (1969) as:

\begin{equation}
  \lambda = \frac{\mid {\bf L}_0 \mid \cdot \sqrt{\mid E_0^{gr} \mid}}
              {G \cdot (M_{gas}+M_{halo})^{5/2}},
   \label{eq:Spin}
\end{equation}

$ {\bf L}_0 $ is the total initial angular momentum and $ E_0^{gr} $ is
the total initial gravitational energy of a protogalaxy. It is to be
noted that for a system in which angular momentum is acquired through
the tidal torque of the surrounding matter, the standard spin parameter
does not exceed $ \lambda \approx 0.11 $ (\cite{SB95}). Moreover, its
typical values range between $ \lambda \approx 0.07^{+0.04}_{-0.05} $,
\eg~ $ 0.02 \le \lambda \le 0.11 $.


\subsection{Dynamical model}

\begin{figure*}[t]

\vsp \vsp \insplt{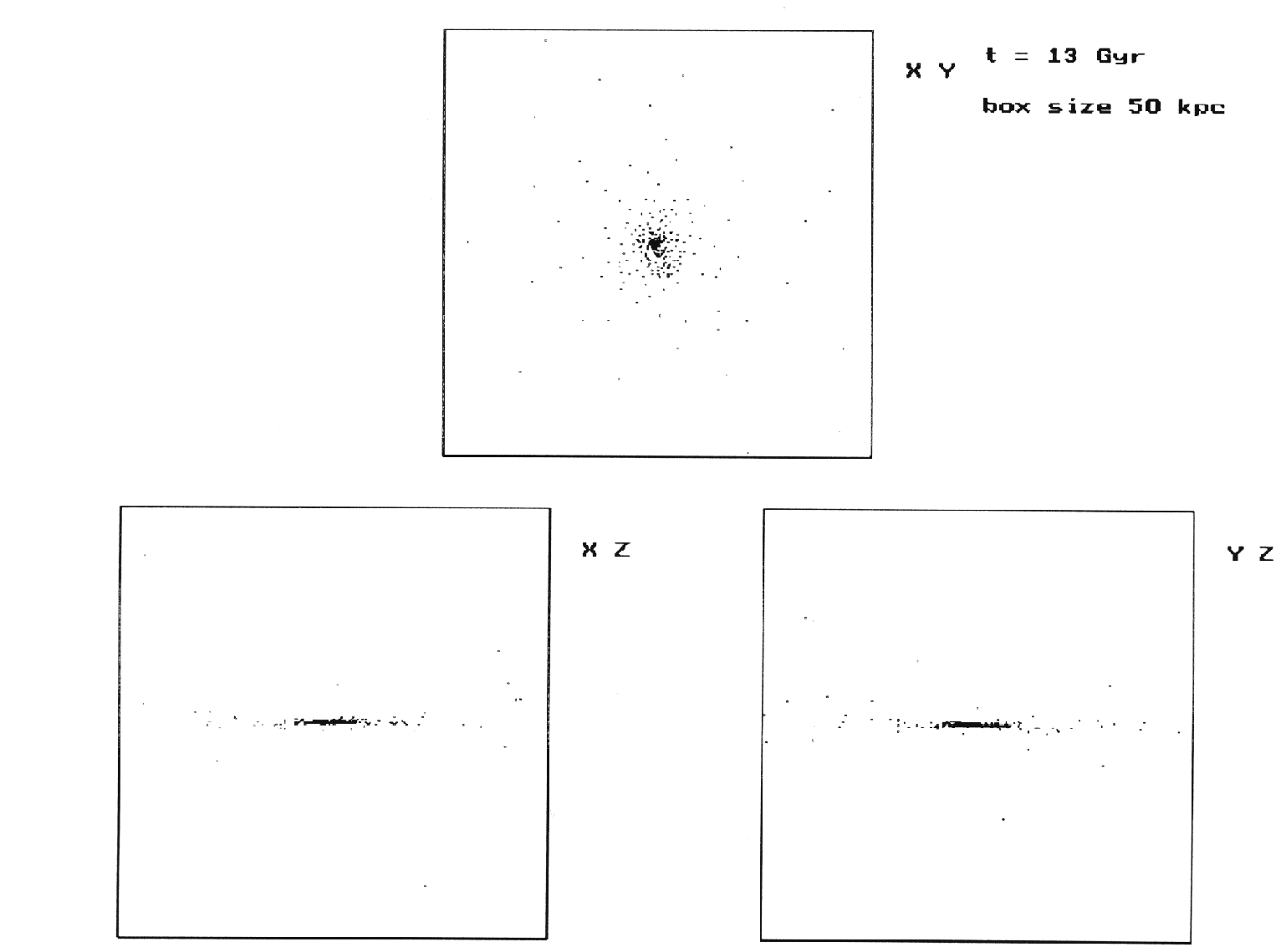}

\caption{The distribution of "gas" particles in the final step}

\label{fig-gas}
\end{figure*}

\begin{figure*}[t]

\vsp \vsp \insplt{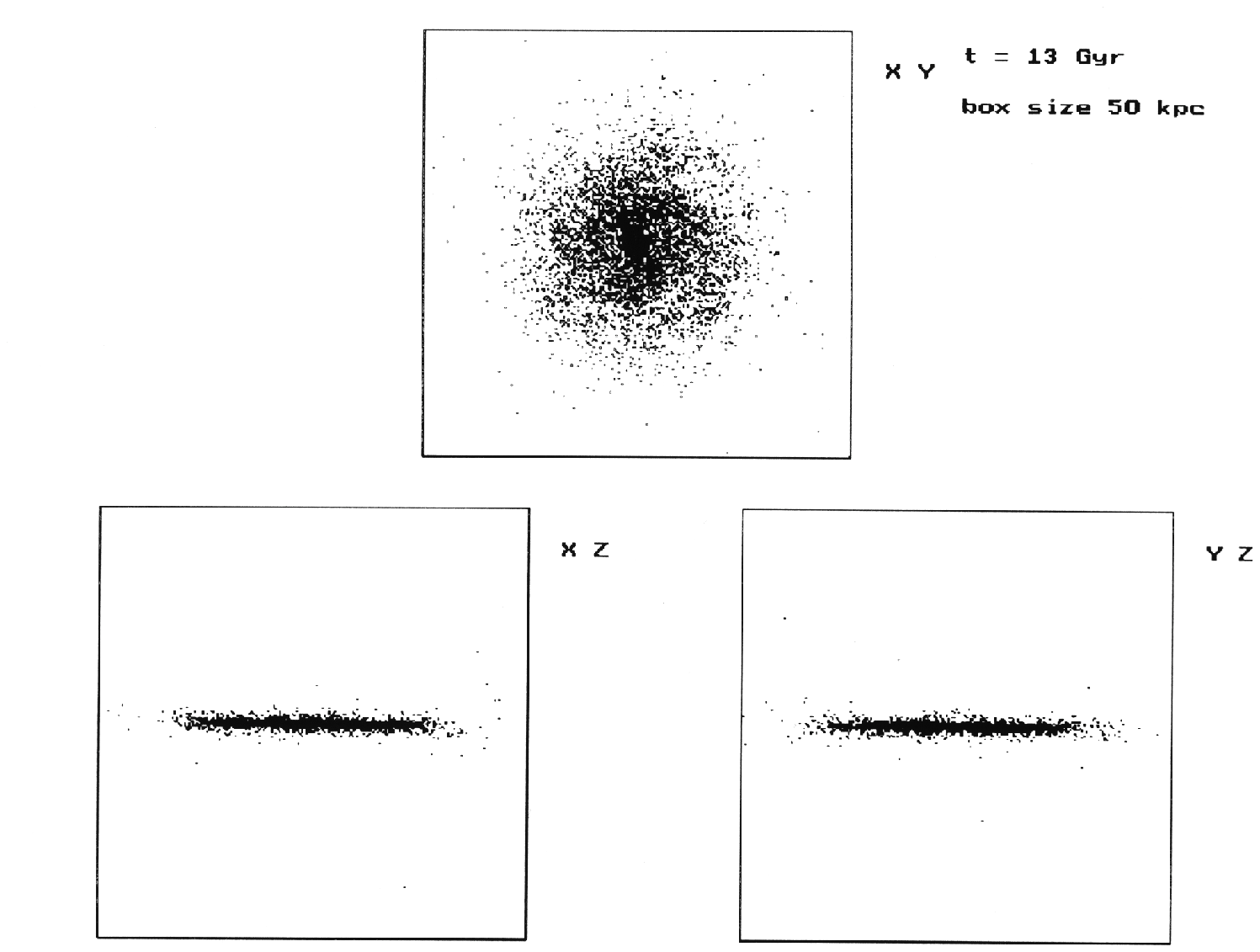}

\caption{The distribution of "star" particles in the final step}

\label{fig-star}
\end{figure*}

In Fig.~\ref{fig-gas}. we present the "XY", "XZ" and "YZ" distributions
of "gas" particles at the final time step ($ t_{evol} \approx 13.0 $
Gyr). The box size is $ 50 $ kpc. In Fig.~\ref{fig-star}. we present
the distributions of "star" particles. The "star" distributions have
dimensions typical of a disk galaxy. The radial extension is
approximately $ 25 - 30 $ kpc. The disk height is around $ 1 - 2 $ kpc.
In the center the "bar -like" structure is developed as a result of
strong initial triaxial structure whole in the plane of the disk we can
see the "spiral - like" distribution of particles, with extended arm
filaments. The "gas" particles are located within central $ 5 - 10 $
kpc.

\begin{figure}[htbp]

\vsp \insertplot{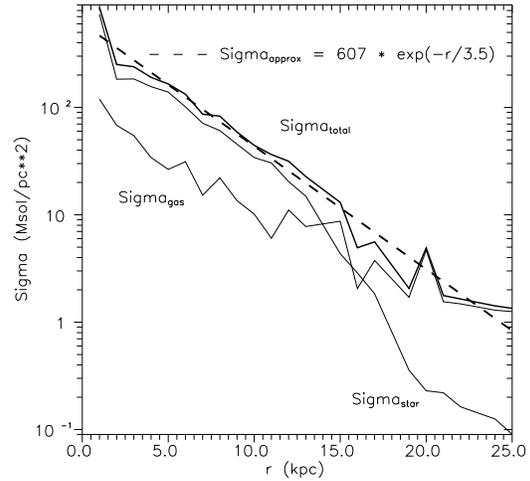}

\caption{$ \sigma_{gas}(r) $,~~~$ \sigma_{*}(r) $ and $ \sigma_{tot}(r)
= \sigma_{gas}(r) + \sigma_{*}(r) $. The column density distribution in
the disk of gas and stars in the final step}

\label{fig-sigma}
\end{figure}

Except for the central region ($ < 2 $ kpc), the gas distribution has a
exponential form with radial scale length $ \approx 2.8 $ kpc. The
column density distributions of gas $ \sigma_{gas}(r) $ and stars $
\sigma_{*}(r) $ are presented in Fig.~\ref{fig-sigma}. The total column
density is defined as: $ \sigma_{tot}(r) = \sigma_{gas}(r) +
\sigma_{*}(r) $. The total column density distribution $
\sigma_{tot}(r) $ is well approximated (in the interval from $ 5 $ kpc
to $ 15 $ kpc) with an exponential profile characterized by a $ \approx
3.5 $ kpc radial scale length. This value is very close to one a
reported recently ($ 3.5 $ kpc) for the radial scale length of the
total disk mass surface density distribution obtained for our Galaxy
(\cite{MCS98b}). The value of $ \sigma_{tot} \approx 55 \; M_\odot $
pc$^{-2}$ near the location of the Sun ($ r \approx 9 $ kpc) is close
to a recent determination of the total density $ 52 \pm 13 \; M_\odot $
pc$^{-2}$ (\cite{MCS98a}).

\begin{figure}[htbp]

\vsp \insertplot{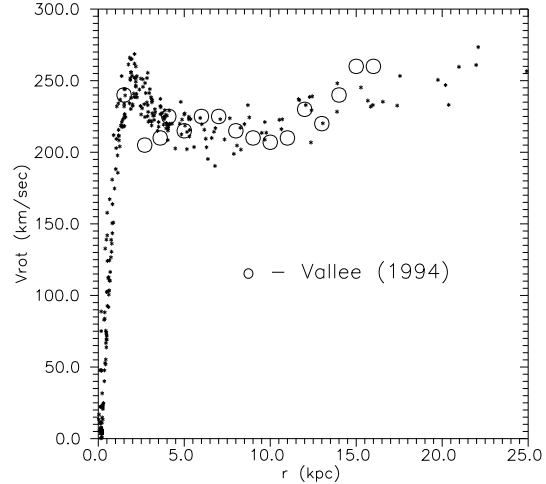}

\caption{$ V_{rot}(r) $. The rotational velocity distribution of gas in
the final step}

\label{fig-v_rot}
\end{figure}

Fig.~\ref{fig-v_rot}. shows both the rotational velocity distribution
of gas $ V_{rot}(r) $ resulting from the modeled disk galaxy
calculation and the rotational curve for our Galaxy (\cite{Va94}), both
of which are very close.

\begin{figure}[htbp]

\vsp \insertplot{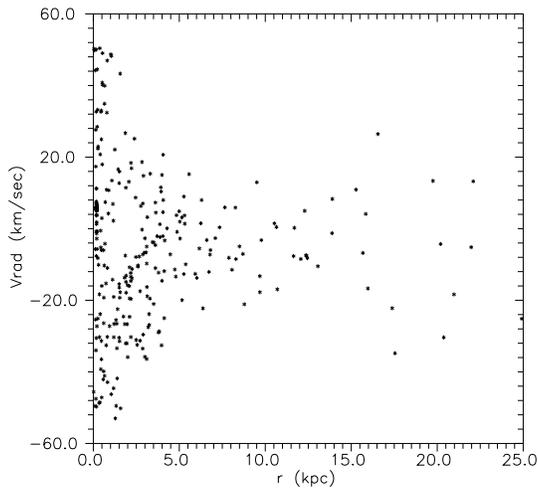}

\caption{$ V_{rad}(r) $. The radial velocity distribution of gas in the
final step}

\label{fig-v_rad}
\end{figure}

\begin{figure}[htbp]

\vsp \insertplot{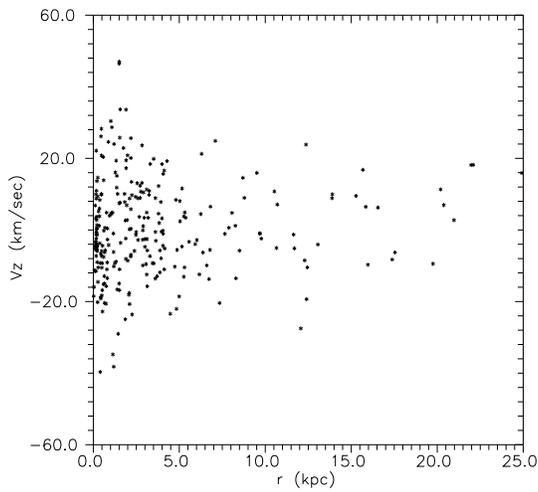}

\caption{$ V_{z}(r) $. The perpendicular to disk normal velocity
distribution of gas in the final step}

\label{fig-v_z}
\end{figure}

The gaseous radial $ V_{rad}(r) $ and normal $ V_{z}(r) $ velocity
distributions are in Fig.~\ref{fig-v_rad}. and Fig.~\ref{fig-v_z}. The
radial velocity dispersion has a maximum value $ \approx 60 $ km/s in
the center, a high value mainly caused by the central strong bar
structure. Near the Sun this dispersion drops down to $ \approx 20 $
km/s. Such radial dispersion is reported in the kinematic study of the
stellar motions in the solar neighborhood (\cite{B98}), while the
normal dispersion is near $ \approx 20 $ km/s in the whole disk. This
value also coincides with the vertical dispersion velocity near the Sun
(\cite{B98}).

\begin{figure}[htbp]

\vsp \insertplot{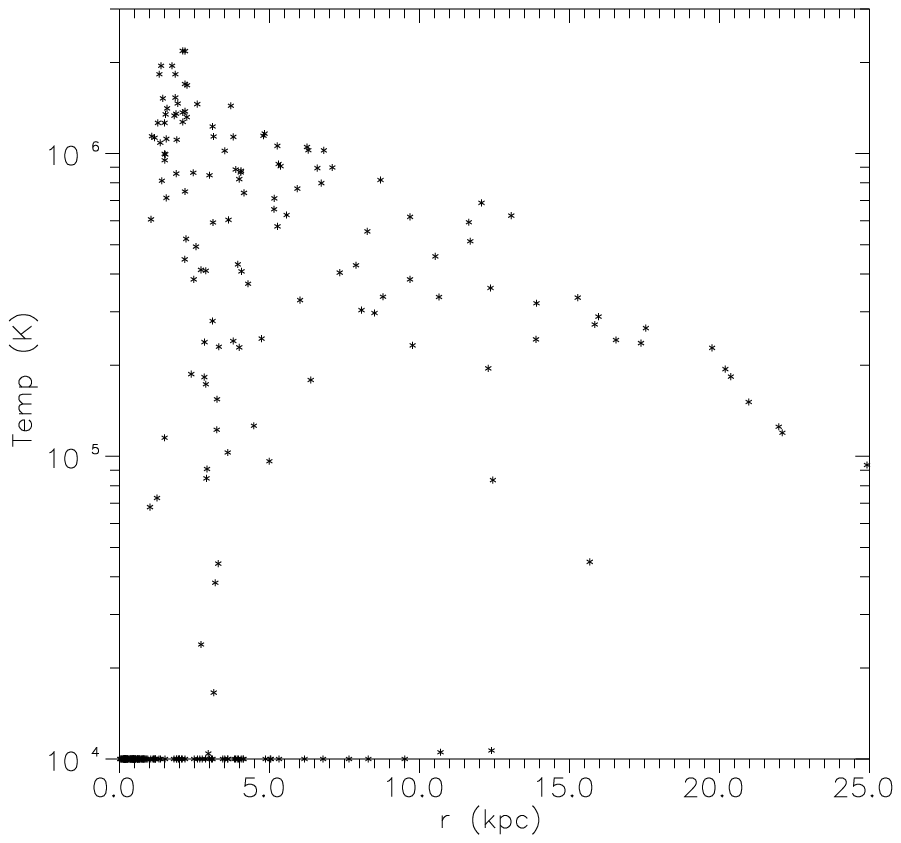}

\caption{$ T(r) $. The temperature distribution of gas in the final
step}

\label{fig-temp}
\end{figure}

We present the temperature distribution of gas $ T(r) $ in
Fig.~\ref{fig-temp}. As seen the distribution of $ T(r) $ has a very
large scatter from $ 10^4 $ K to $ 10^6 $ K. In our calculation we set
the cutoff temperature for the cooling function at $ 10^4 $ K, the gas
can't cool to lower temperatures.

The modeled process of SNII explosions injects to a great amount of
thermal energy into the gas and generates a very large temperature
scatter, also typical of our Galaxy's ISM. At each point even with
crude numerical approximations a good fit can be reached for all
dynamical and thermal distributions of gas and stars in a typical disk
galaxy like our Galaxy.

\subsection{Chemical characteristics}

\begin{figure}[htbp]

\vsp \insertplot{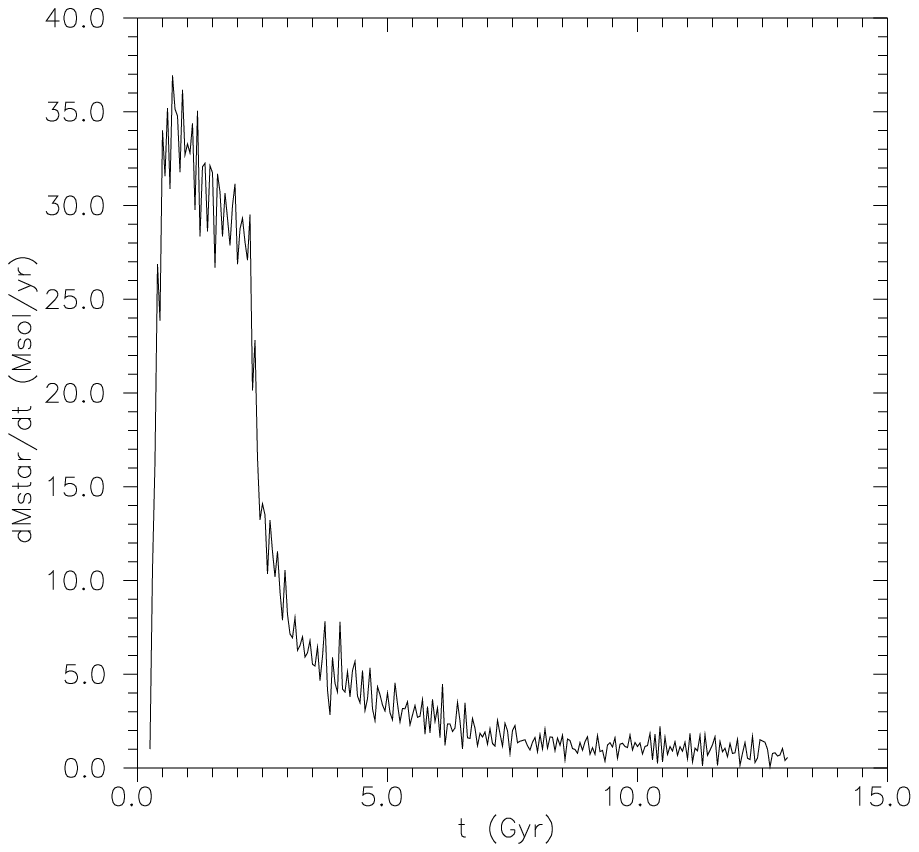}

\caption{$ SFR(t) = dM_{*}(t)/dt $. The time evolution of the SFR in
galaxy}

\label{fig-dm}
\end{figure}

Fig.~\ref{fig-dm}. shows the time evolution of the SFR in galaxy $
SFR(t) = dM_{*}(t)/dt $. Approximately $ 90 \% $ of gas is converted
into stars at the end of calculation. The most intensive SF burst
happened in the first $ \approx 1 $ Gyr, with a maximal SFR $ \approx
35 \; M_\odot/yr $. After $ \approx 1.5 - 2 $ Gyr the SFR is decreases
like an "exponential function" until it has a value $ \approx 1 \;
M_\odot/yr $ at the end of the simulation. To check the SF and chemical
enrichment algorithm in our SPH code, we use the chemical
characteristics of the disk in the "solar" cylinder ($ 8 $ kpc $ < \; r
\; < \; 10 $ kpc).

\begin{figure}[htbp]

\vsp \insertplot{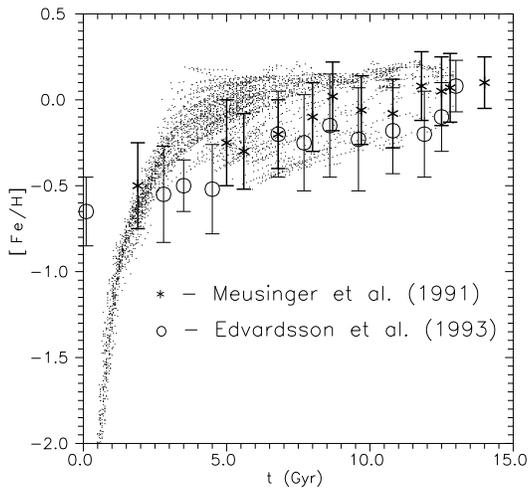}

\caption{[Fe/H]$(t)$. The age metallicity relation of the "star"
particles in the "solar" cylinder ($ 8 $ kpc $ < \; r \; < \; 10 $
kpc)}

\label{fig-chem_1}
\end{figure}

\begin{figure}[htbp]

\vsp \insertplot{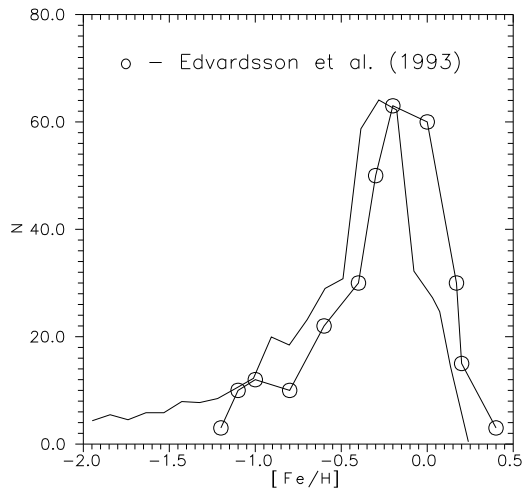}

\caption{$N_{*}($[Fe/H]$)$. The metallicity distribution of the "star"
particles in the "solar" cylinder ($ 8 $ kpc $ < \; r \; < \; 10 $
kpc)}

\label{fig-chem_2}
\end{figure}

The age--metallicity relation of the "star" particles in the "solar"
cylinder, [Fe/H]$(t)$, is shown in Fig.~\ref{fig-chem_1}., with
observational data taken from
Meusinger \etal~ (1991) and Edvardsson \etal~ (1993), while in
Fig.~\ref{fig-chem_2}. we presented the metallicity distribution of the
"star" particles in the "solar" cylinder $N_{*}($[Fe/H]$)$. The model
data are scaled to the observed number of stars (\cite{EAG93}). In
Fig.~\ref{fig-chem_1}. each model point represents the separate "star"
particle. The mass of each "star" particle is different (from $ \sim
10^4 \; M_\odot $ up to $ \sim 2.5 \; 10^6 \; M_\odot $), because the
star formation efficiency  - $ \epsilon $ is different in each star
forming region. The model point is systematically higher than the
observations (especially near the $ t \approx 5 $ Gyr), but if we also
analyze the mass of each "star" particle we see that the more massive
particles systematically show lower metallicity than the observations.
If one divides the metallicity to equal zone and calculate the sum of
the mass in each metallicity zone in Fig.~\ref{fig-chem_1}. we get the
results in Fig.~\ref{fig-chem_2}. and in this figure we see what the
model mass distribution has shifted to the lower metallicities.

\begin{figure}[htbp]

\vsp \insertplot{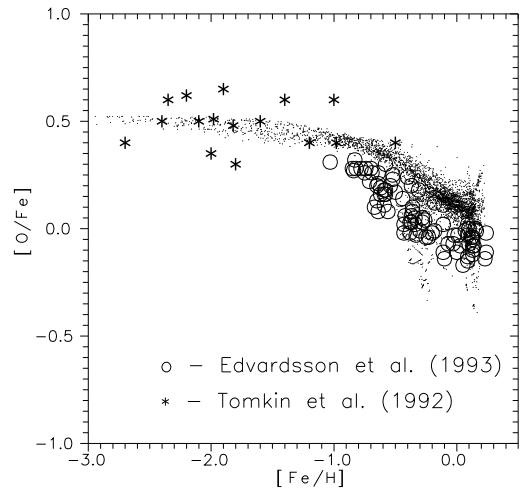}

\caption{The [O/Fe] vs. [Fe/H] distribution of the "star" particles in
the "solar" cylinder ($ 8 $ kpc $ < \; r \; < \; 10 $ kpc)}

\label{fig-chem_3}
\end{figure}

The [O/Fe] vs. [Fe/H] distribution of the "star" particles in the
"solar" cylinder one found in Fig.~\ref{fig-chem_3}. In this figure we
also present the observational data from
Edvardsson \etal~ (1993) and Tomkin \etal~ (1992). All these model
distributions are in good agreement, not only with presented
observational data, but also with other data collected from
Portinari \etal~ (1997).

\begin{figure}[htbp]

\vsp \insertplot{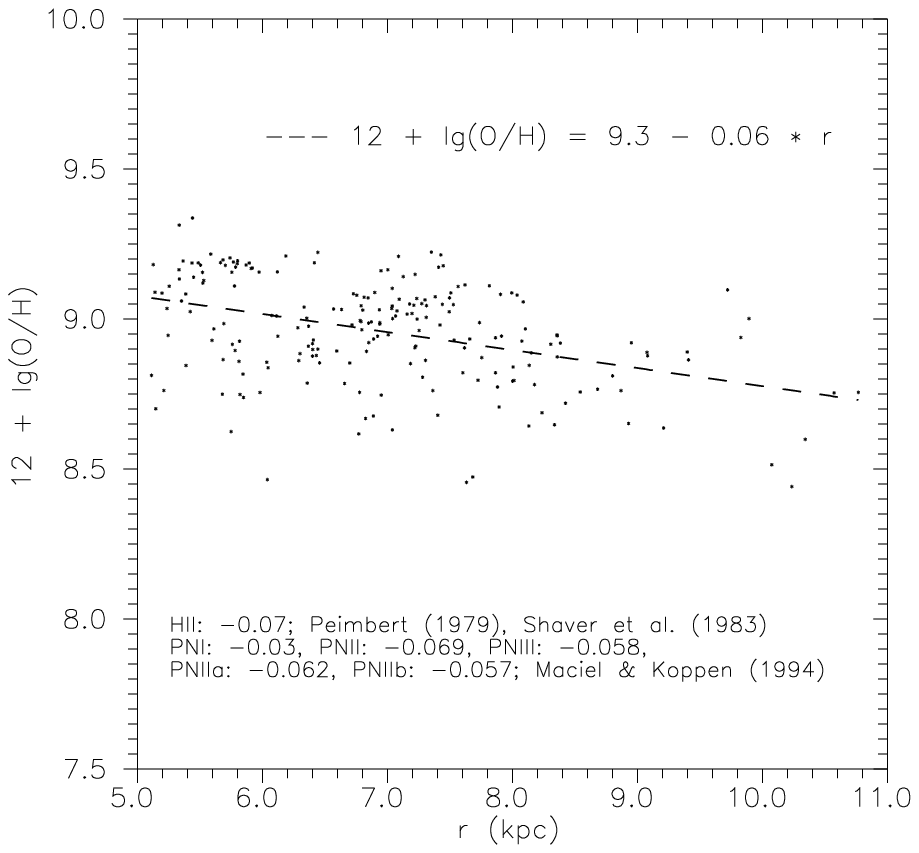}

\caption{[O/H]$(r)$. The [O/H] radial distribution}

\label{fig-grad-z}
\end{figure}

The [O/H] radial distribution [O/H]$(r)$ is shown in
Fig.~\ref{fig-grad-z}. The approximation presented in the figure is
obtained by a least - squares linear fit. At distances $ 5 $ kpc $ < \;
r \; < \; 11 $ kpc the models radial abundance gradient is $ -0.06 $
dex/kpc. In the literature we found different values of this gradient
defined in objects of different types. From observations of HII regions
(\cite{P79}; \cite{SMNDP83}) we obtained oxygen radial gradient $ -0.07
$ dex/kpc. From observations of PN of different types (\cite{MK94}) we
obtained the values: $ -0.03 $ dex/kpc for PNI, $ -0.069 $ dex/kpc for
PNII, $ -0.058 $ dex/kpc for PNIII, $ -0.062 $ dex/kpc for PNIIa, and $
-0.057 $ dex/kpc for PNIIb. All this agrees well with the oxygen radial
gradient in our Galaxy.

\subsection{Conclusion}

This simple model provides a reasonable, self--consistent picture of
the processes of galaxy formation and star formation in the galaxy. The
dynamical and chemical evolution of the modeled disk--like galaxy is
coincident with observations for our own Galaxy. The results of our
modeling give a good base for a wide use of the proposed SF and
chemical enrichment algorithm in other SPH simulations.


{\em Acknowledgments.} The author is grateful to S.G. Kravchuk, L.S.
Pilyugin and Yu.I. Izotov for fruitful discussions during the
preparation of this work. It is a pleasure to thank Pavel Kroupa and
Christian Boily for comments on an earlier version of this paper. The
author also thanks the anonymous referee for a helpful referee's report
which resulted in a significantly improved version of this paper. This
research was supported by a grant from the American Astronomical
Society.



\end{document}